\documentclass[12pt,preprint]{aastex}

\slugcomment{Accepted to the Astronomical Journal}

\shorttitle{H$_3^+$ Emission from Hot Jupiters}
\shortauthors{Shkolnik et al.}

\begin{document}


\title{No Detectable H$_3^+$ Emission from the Atmospheres of Hot Jupiters}


\author{Evgenya Shkolnik\altaffilmark{1,2,3}}
\affil{Institute for Astronomy, University of Hawaii at Manoa, 2680 Woodlawn Drive, Honolulu, HI 96822}
\email{shkolnik@hawaii.edu}

\author{Eric Gaidos\altaffilmark{1,3}}
\affil{Department of Geology \& Geophysics, University of Hawaii at Manoa}

\and
\author{Nick Moskovitz\altaffilmark{1}}
\affil{Institute for Astronomy, University of Hawaii at Manoa}


\altaffiltext{1}{Visiting Astronomer at the Infrared Telescope Facility, which is operated by the University of Hawaii under Cooperative Agreement No. NCC 5-538 with the National Aeronautics and Space Administration, Office of Space Science, Planetary Astronomy Program.}
\altaffiltext{2}{NRC Research Associate}
\altaffiltext{3}{Partially supported by the National Aeronautics and Space Administration through the NASA Astrobiology Institute under Cooperative Agreement No. NNA04CC08A issued through the Office of Space Science.}


\begin{abstract}
H$_3^+$ emission is the dominant cooling mechanism in Jupiter's thermosphere and a useful probe of temperature and ion densities. The H$_3^+$ ion is predicted to form in the thermospheres of close-in `hot Jupiters' where its emission would be a significant factor in the thermal energy budget, affecting temperature and the rate of hydrogen escape from the exosphere. Hot Jupiters are predicted to have up to 10$^5$ times Jupiter's H$_3^+$ emission because they experience extreme stellar irradiation and enhanced interactions may occur between the planetary magnetosphere and the stellar wind. Direct (but unresolved) detection of an extrasolar planet, or the establishment of useful upper limits, may be possible because a small but significant fraction of the total energy received by the planet is re-radiated in a few narrow lines of H$_3^+$ within which the flux from the star is limited. 

We present the observing strategy and results of our search for emission from the Q(1,0) transition of H$_3^+$ (3.953~$\mu$m) from extrasolar planets orbiting six late-type dwarfs using CSHELL, the high-resolution echelle spectrograph on NASA's Infrared Telescope Facility (IRTF). We exploited the time-dependent Doppler shift of the planet, which can be as large as 150 km s$^{-1}$, by differencing spectra between nights, thereby removing the stellar photospheric signal and telluric lines. We set limits on the H$_3^+$ emission from each of these systems and compare them with models in the literature.  Ideal candidates for future searches are intrinsically faint stars, such as M dwarfs, at very close distances. 

\end{abstract}





\section{Introduction}

Twenty percent of the 180 known extrasolar planets orbit within 0.1~AU of their parent star and have masses comparable to Jupiter's. The successful radial velocity technique (e.g., Mayor \& Queloz 1995; Marcy \& Butler 1996) provides the Keplerian parameters for these `hot Jupiters': minimum mass (M$_p$sin$i$), orbital period (P$_{orb}$), semi-major axis ($a$), and eccentricity ($e$).  For the very small number of these that transit their star, the transit lightcurves supply the planet radius (R$_p$), orbital inclination ($i$), and true mass (M$_p$; e.g.~Charbonneau et al.~2000, 2006), and in the rare case of a bright host star, spectroscopy of the transit event can allow one to measure or constrain elemental and molecular abundance in the planet's atmosphere.
Our knowledge of a hot Jupiter's atmospheric composition is limited to the case of HD~209458~b, whose parent star is bright enough (V=7.65) for `transit spectrophotometry.' This method provided the first detection of an extrasolar planetary atmosphere through observations of Na I absorption by the planet's troposphere (Charbonneau et al.~2002).


Vidal-Madjar et al.~(2003, 2004), using transit spectroscopy, reported a detection of ultraviolet absorption by H I, O I and C II in the upper atmosphere of HD~209458~b out to several planetary radii.\footnote{These observations were acquired with the Space Telescope Imaging Spectrograph (STIS) aboard the Hubble Space Telescope. Because STIS is no longer operational, the observations cannot be presently confirmed.} They inferred that the observed 15\% Ly$\alpha$ absorption is taking place beyond the planet's Roche limit, and thus that hydrogen was escaping at a rate of $\sim$$10^{10}$ g~s$^{-1}$.
This observation motivated several models of the upper atmosphere of hot Jupiters that experience the influence of the parent star's extreme ultraviolet (EUV) flux on temperature and escape rates. Rapid hydrodynamic escape of hydrogen would potentially affect the planet's structure, evolution, and shed light on the mass-radius relationship of the known transiting planets (Pont et al.~2005). The first such calculation was presented by Lammer et al.~(2003), who based their model on Watson et al.'s (1981) hydrodynamic treatment in which upper atmospheric temperature and mass loss are coupled. Lammer et al. pointed out that even though the effective temperatures of hot Jupiters are low enough for the planets to be stable against mass loss, UV-heated exosphere temperatures are much higher, promoting escape. They calculated a H escape of  $\sim$10$^{12}$~g~s$^{-1}$, 9 orders of magnitude greater than the Jeans escape at the planet's effective temperature (Sasselov 2003). Baraffe et al.~(2004) applied this work to evolutionary calculations where they unite irradiated planet atmospheres with interior structures. They show that for a given orbital distance there is a critical mass  below which the planet will undergo rapid runaway evaporation, and though extremely improbable, HD~209458~b might be in such a fleeting phase. 

Using a more detailed model, Yelle (2004) calculated that the energy-limited atmospheric escape rate is approximately proportional to the stellar EUV flux and for HD 209458 b, is 20 times less than the Lammer et al.~value. This difference is ascribed to the dependence of escape rates on the thermal structure of the upper atmosphere and the previously neglected balance between heating by EUV radiation and cooling by escape of H and radiation by molecules in the homopause at infrared wavelengths. Yelle (2004) showed that the near IR emission by H$_3^+$ is an important cooling mechanism in the upper atmosphere of hot Jupiters just as it is in Jupiter's hot thermosphere.

In the following section, we present the predictions of H$_3^+$  emission from a hot Jupiter which encouraged this observing program. In addition to being the first detection of H$_3^+$ in emission outside of the Solar System,\footnote{H$_3^+$ has been observed in absorption in the interstellar medium (e.g.~Goto et al.~2002). Brittain \& Rettig (2002) announced a detection of H$_3^+$ emission from the protoplanetary disk around HD 141569, but more detailed observations by Goto et al.~(2005) failed to confirm the result.} direct planetary detection would offer a new view into the physical properties of hot Jupiter thermospheres, applying needed constraints to models of their thermal structure. We present our search for the Q(1,0) transition of H$_3^+$ ($\lambda$ = 3.9530 $\mu$m or 2529.72 cm$^{-1}$) from the upper atmospheres of six close-in extrasolar planets. This wavelength is in the L$^\prime$ passband, where the parent star, relative to optical wavelengths, is dim making the contrast with planetary emission more favorable. We capitalized on the time-dependent Doppler shift of the planet, which can be as large as 150 km s$^{-1}$ (or $\Delta\lambda = 0.0020 \mu$m at 4$\micron$) between two nights. We differenced spectra to remove the stellar photospheric signal and telluric absorption lines, isolating any residual H$_3^+$ emission which would stand out as a peak and a velocity shifted dip in the final residual. Section~\ref{obs} outlines our observations and data reduction, while the analysis and resulting luminosity limits are presented in Section~\ref{results}. We summarize our results and discuss the requirements of future experiments in Section~\ref{summary}.

\section{H$_3^+$ emission from hot Jupiters\label{predictions}}

Jovian H$_3^+$ has an equatorial column density of $\sim$10$^{15}$~m$^{-2}$ and is $\sim$100 times more abundant at auroral latitudes (Lam et al.~1997). It has been successfully used as a diagnostic of the thermal and chemical state of Jupiter's thermosphere and ionosphere (Stallard et al.~2002).\footnote{The solar EUV radiation and magnetospheric charged particles that are absorbed by Jupiter's thermosphere dissociate or ionize the thermospheric molecules creating a coincident ionosphere (Yelle \& Miller 2004).} H$_3^+$ has a strong ro-vibrational spectrum emitting at 3$-$4 $\mu$m, corresponding to excitation temperatures of $\sim$1000 and 750~K, and thereby effectively sampling the temperature range of Jupiter's upper atmosphere (Stallard 2001). 

Several groups modeled the formation of H$_3^+$ in giant planet ionospheres (e.g.~Kim et al.~1992, Achilleos et al.~1998). A chain reaction starting with the ionization of H$_2$ forms the molecular ion in the thermosphere. At high latitudes, this is achieved by collisions with energetic electrons funneled down along magnetic field lines: 

\centerline{H$_2$ + e$^* \rightarrow$ H$_2^+ +$ e$^-$ + e$^-$,}

\noindent and more globally, through stellar EUV radiation,

\centerline{H$_2$ + h$\nu \rightarrow$ H$_2^+ +$ e$^-$.}

\noindent An exothermic reaction quickly converts the ion to H$_3^+$,

\centerline{H$_2^+$ + H$_2 \rightarrow$ H$_3^+$ + H.}

\noindent H$_3^+$ has a minimum lifetime of about 10 s (Achilleos et al.~1998) and is destroyed by dissociative recombination:

\centerline{H$_3^+$ + e$^-$ $\rightarrow$ H$_2^+$ + H + e$^-$}

\centerline{or  H$_3^+$ + e$^-$ $\rightarrow$ H + H + H.}

Jupiter emits $\sim$10$^{12}$ W in the emission lines of H$_3^+$  (Lam et al.~1997). At a distance of 10~pc, the resulting flux of 8$\times$10$^{-25}$~W~m$^{-2}$ (=2$\times$10$^{-20}$~W~m$^{-2}$~$\mu$m$^{-1}$ or 0.03 $\mu$Jy if placed in an unresolved line at spectral resolution of 36,000) would be undetectable with today's telescope technologies. A hot Jupiter ($a \sim$ 0.05 AU) is 100 times closer to its parent star and experiences 10$^4$ times the EUV flux and enhanced magnetic interaction with the stellar magnetosphere (Shkolnik et al.~2005). Direct detection of a transiting or even non-transiting hot Jupiter atmosphere is possible because a significant fraction of this additional energy is re-radiated by narrow lines of molecular collants, including H$_3^+$. As predicted by Yelle (2004) for hot Jupiters and measured by Stallard et al.~(2002) in our own Jupiter (see below), H$_3^+$ emission is the dominant coolant of the planet's thermosphere.  A measurement of H$_3^+$ emission would provide needed information about temperature and ion density in the thermosphere of a hot Jupiter.

Miller et al.~(2000) used the Jupiter Ionosphere Model (JIM; Achilleos et al.~1998), a three-dimensional, fully coupled model of the Jovian thermosphere/ionosphere, to calculate that a Jupiter orbiting the Sun at 0.05 AU would have an H$_3^+$ column density (due to irradiation alone) at the sub-solar point of 10$^{18}$~m$^{-2}$, or 1000 times that of Jupiter. The predicted emission 
is  $\sim$10$^{17}$~W, or a few $\times$10$^{-21}$ W~m$^{-2}$ if the planet is 15 pc away. They consider this a lower limit because their model does not take into consideration an extended atmosphere as observed for HD~209458~b, a deeper atmosphere for planets that are more massive than Jupiter, any increased ionization due to the strong stellar wind, or interactions between the magnetospheres of the planet and the star.

Yelle (2004) presented a 1-dimensional aeronomical calculation representing a global-average of the atmospheres of hot Jupiters with $a$ = 0.01, 0.05, and 0.1 AU.  The model shows that the planets' thermospheres are heated beyond 10$^4$ K by the EUV radiation from the solar-type star. The lower thermospheres, between 1 and 1.1 R$_p$, are primarily cooled by H$_3^+$ radiative emission, despite the fact that the calculated H$_3^+$ densities are approximately a factor of 10 lower than in Jupiter. This is because the increased photoionization both creates H$_2^+$, which is quickly converted to H$_3^+$, and generates the electrons that destroy the H$_3^+$.  
Cooling rates vary from 6$\times10^{-8}$ ergs cm$^{-3}$ s$^{-1}$ at 0.1 AU to 3$\times10^{-6}$ ergs cm$^{-3}$ s$^{-1}$  at 0.01 AU. For the reference case, modeled after HD~209458~b ($a$ = 0.05 AU, R$_p$ = 1.4R$_J$), Yelle predicted an H$_3^+$ luminosity of 1.0$\times$10$^{16}$~W, an order of magnitude less than Miller et al.'s prediction. 

Neither model includes the effects of the stellar gravitational force, stellar radiation pressure, or a planetary magnetic field.  The first two become important only beyond  $\sim$3~R$_p$, however a magnetic field would make a direct contribution to the ionosphere. Though we do not yet know how ubiquitous magnetic fields are among extrasolar planets, Shkolnik et al.~(2003, 2005) presented evidence suggesting strong fields (B $\geq$ B$_J$ = 4.3 G) on two hot Jupiters, HD~179949~b and $\upsilon$~And~b. 
A planetary field would trap ions, limiting ion escape, and cause the precipitation of electrons and ions along magnetic field lines, producing polar enhancements of H$_3^+$ on the close-in planets similar to those observed on our Jupiter.  This might increase the H$_3^+$ emission from a hot Jupiter, potentially by orders of magnitude beyond the predictions.

\section{Observations \& data reduction\label{obs}}

Our six stellar targets have a broad range of spectral types (F7$-$M2.5) and L$^\prime$ magnitudes (2.8$-$7.8).  Each star hosts either a hot Jupiter- or a hot Neptune-mass inner planet with an orbital period of $\lesssim$7 days. Given the uncertainties in the predictions, there was a possibility that detectable H$_3^+$ emission might have come from planets farther away from their central star.  For this reason, we included three stellar systems that have one or more additional outer planets. Table~\ref{planets} lists each planet's minimum mass, orbital period, and semi-major axis, while the stellar properties are given in Table~\ref{stars}.

We secured two nights of L$^{\prime}$-band spectra of the six stars on the 3-m NASA Infrared Telescope Facility's (IRTF) Cryogenic Near-IR Facility Spectrograph (CSHELL) in its high-resolution echelle mode during 2005 February 15$-$17 and August 26 \& 28. (August 27 was lost to poor weather.)  A  0.5$^{\prime\prime}\times30^{\prime\prime}$ slit gave a spectral resolution of 36,000 with an average resolution element of 2.70 detector pixels (8.3 km~s$^{-1}$). Narrow-band circular variable filters (CVF) isolate a single order within a spectral range of 1$-$5$\mu$m.  The 256$\times$160 SBRC InSb detector (Tokunaga et al.~1990; Greene et al.~1993) intercepts a narrow wavelength range equal to 1/400 of the central wavelength, in this case, 3.948$-$3.958 $\mu$m.

\begin{figure}
\epsscale{.80}
\plotone{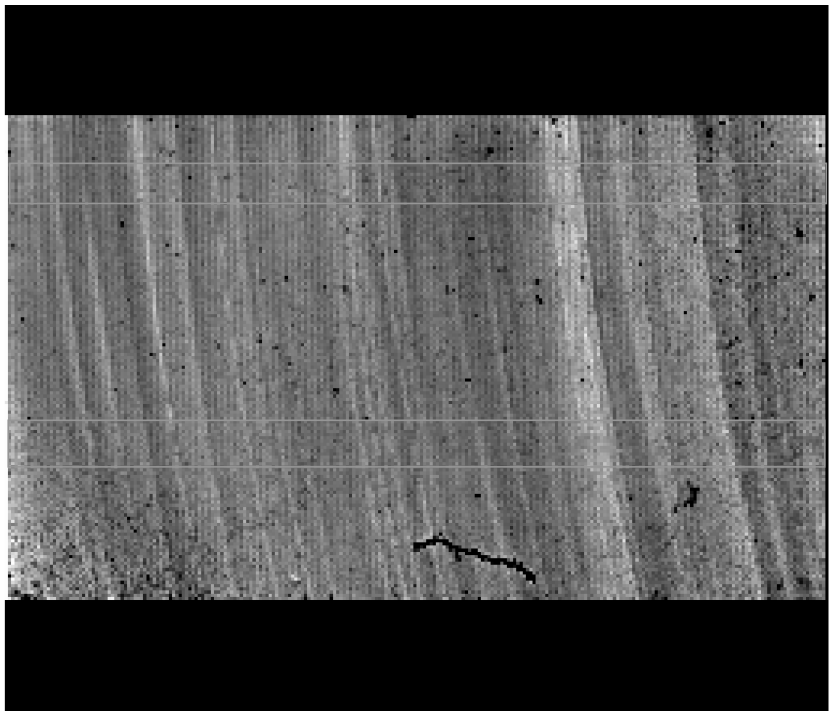}
\caption{Image of the normalized flat field. The lower and upper strips represent the locations of the A and B beams, respectively. The fringing that induces variations in the two beams is evident.\label{flat_n}}
\end{figure}

We nodded the star along the slit with a beam pattern of A-B-B-A to ensure accurate sky subtraction. The detector's response becomes non-linear beyond 3000 DN (33,000 e$^-$), which forced us to keep all exposures no longer than 70$-$120 seconds.  The total integration times and resulting S/N for each star are listed in Table~\ref{stars}.
The data reduction was performed with standard IRAF\footnote{IRAF is distributed by the National Optical Astronomy Observatories, which are operated by the Association of Universities for Research in Astronomy, Inc., under cooperative agreement with the National Science Foundation.} routines. We first applied a bad pixel mask to correct for $\sim$250 ``noisy" (non-linear) pixels, and eliminated cosmic ray hits with the {\it cosmicray}~procedure. We produced a nightly mean dark-subtracted, normalized 2-D flat field to flatten each [A$-$B] and [B$-$A] frame.  Movement of the CVF wheel resulted in asymmetric fringing and imperfect flat-fielding (Figure~\ref{flat_n}, J.~Rayner, personal communication). This produced inconsistencies between the A and B beams and cannot be avoided with CSHELL since one needs to move the CVF wheel to acquire a star, and again to take a flat field exposure. We therefore treated each beam separately throughout the data reduction and analysis, combining only the final A- and B-beam differenced spectra (residuals, where the fringing effects are removed) for increased S/N. 

We extracted a 1-D spectrum from each sky-subtracted, flat-fielded 2-D image and calibrated wavelength with a krypton arc lamp spectrum. We took arc exposures before and after each stellar target. Due to the very small spectral range, it is rare for an arc line to fall within this span.  It was necessary to observe arc lines at the same grating position as used for H$_3^+$ line, move the CVF filter to select three different grating orders, and extrapolate to the desired wavelengths using the grating law (relating orders, angles, and wavelengths; see CSHELL manual).  One-dimensional arc spectra were extracted for each of the two beams since the curvature in the arc lines across the detector produced a small difference in the wavelength solutions.  The final precision of the wavelength solution is 2.5$\times$10$^-5$~$\micron$. 

We were unable to remove the N$_2$O telluric absorption lines due to the poor S/N of our hot standard star spectra. This did not pose a problem for most of the stellar targets when comparing two nights of data because the atmospheric N$_2$O lines were stable. However, we could not apply heliocentric velocity corrections with the telluric lines in the spectra, or ``shift \& add" the individual exposures to accommodate for potential broadening of the H$_3^+$ line due to the planet's orbital motion within the nightly span of observations.  These velocities are for the most part negligible at this spectral resolution.  The heliocentric (and barycentric) velocities differ from night to night by less than 3 km~s$^{-1}$.  The broadening of any planetary emission due to orbital motion in a couple of hours is usually $\lesssim$8~km~s$^{-1}$, or one resolution element.  Broadening due to the planet's rotation is also insignificant since these hot Jupiters are likely tidally locked to their stars, resulting in rotation rates of $\lesssim$1~km~s$^{-1}$.

In the end, spectra of each target star on a given night were combined to produce the final A- and B-beam spectra.  The sought-after H$_3^+$ emission would be greatly Doppler shifted (150 km~s$^{-1}$) and stand out as a peak and a velocity shifted dip in the mean residual (=$\frac{1}{2}$[(A$_{Night 1}$ $-$ A$_{Night 2}$) + (B$_{Night 1}$ $-$ B$_{Night 2}$)]).

\section{Results\label{results}}

\begin{figure}
\includegraphics[width=\textwidth]{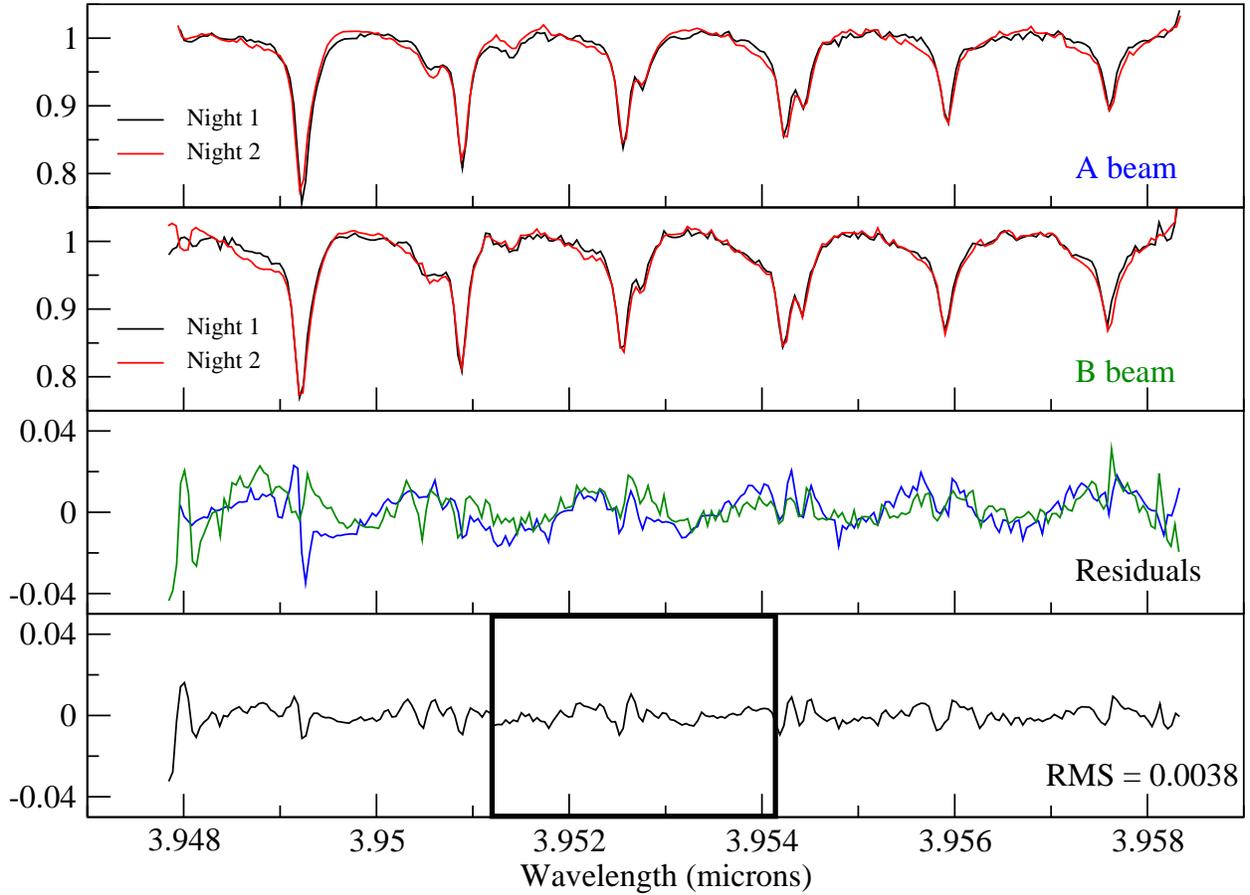}
\caption{L$^\prime$ spectra of $\upsilon$~And centered on the H$_3^+$ 3.953-$\mu$m line.  Two nights of CSHELL's A-beam and B-beam are plotted in the first two panels. The y-axes are normalized intensity. The third panel shows the residuals of the two beams. The periodic structure in the beams reflects the change in the N$_2$O telluric lines  and is removed in the bottom panel which plots the mean residual. The boxed region spans the wavelength range where the Doppler-shifted H$_3^+$ line would be and has an RMS of 0.38\%.\label{upsAnd}}
\end{figure}

L$^\prime$-band spectra of $\upsilon$~And, one of our brightest targets, are displayed in Figure~\ref{upsAnd}. We plot the normalized A- and B-beam spectra taken on 2005 August 26 and 28 in the top two panels with their corresponding residuals in the third panel.  
A Fourier analysis revealed a peak frequency of 575~$\mu$m$^{-1}$ (0.00244 cycles~pixel$^{-1}$) in both the $\upsilon$~And spectra and their residuals. We removed this frequency  from the residuals as it is clearly due to small variability in telluric absorption.  The final mean residual is shown in the fourth panel of Figure~\ref{upsAnd}.  The boxed region spans the wavelength range where the Doppler-shifted H$_3^+$  emission would appear if detected. The RMS within this region is 0.38\% of the stellar photosphere signal. The residual RMS is converted into an H$_3^+$ luminosity limit using the star's L$^\prime$ magnitude and distance.  We derived the L$^\prime$ magnitudes from the K magnitudes of the Two Micron All Sky Survey (2MASS, Skrutskie et al.~2006) and intrinsic colors from Bessell \& Brett (1988). For $\upsilon$~And, the RMS translates into a luminosity limit of 1.3$\times$10$^{18}$ W. The RMS of the intranight residuals for the $\upsilon$~And spectra is 0.2\%. This demonstrates the level of stability of the spectrograph on short timescales as well as the reliability of the data reduction and analysis. 

\begin{figure}
\includegraphics[width=\textwidth]{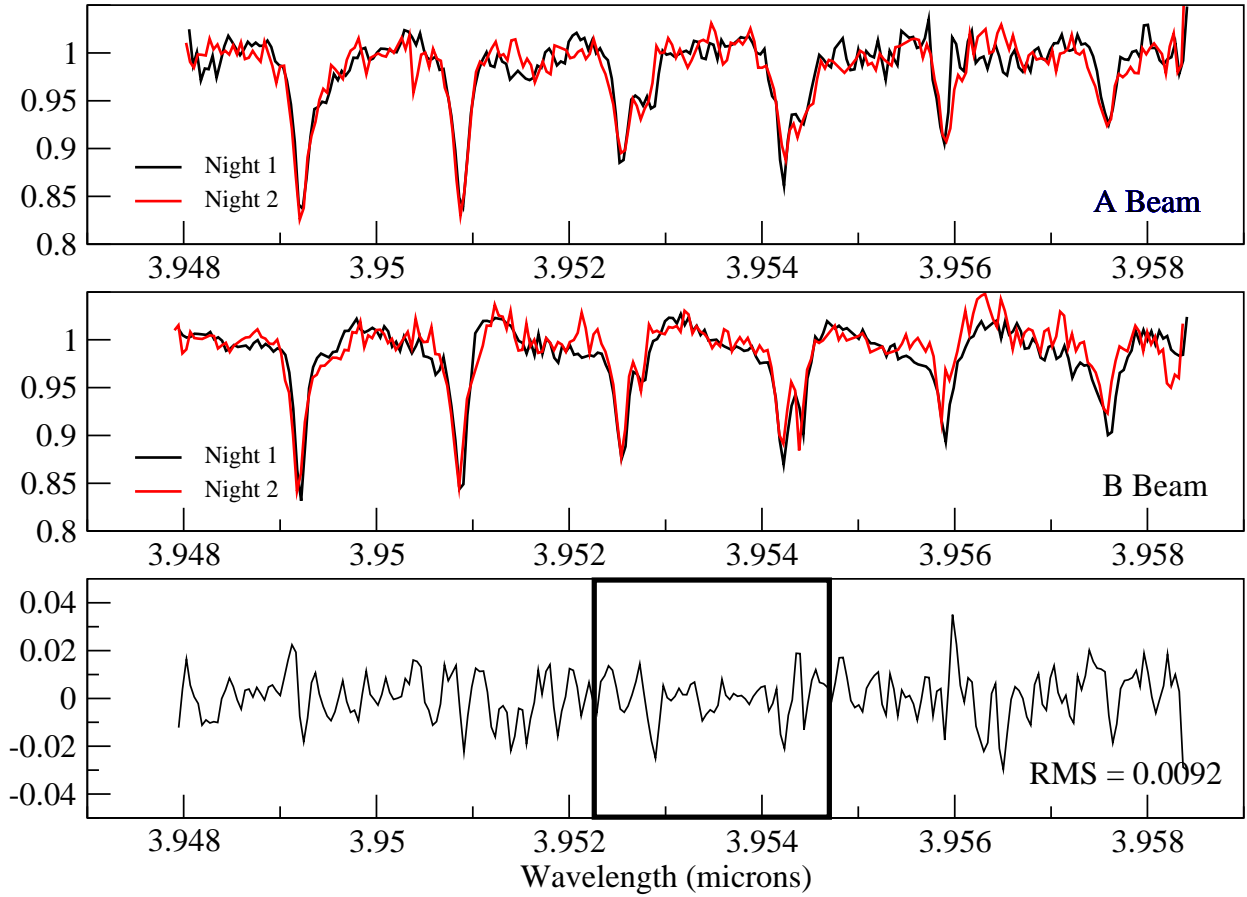}
\caption{L$^\prime$ spectra of $\tau$ Boo centered on the H$_3^+$ 3.953-$\mu$m line.  Two nights of CSHELL's A-beam and B-beam are plotted in the first two panels. The y-axes are normalized intensity. The third panel shows the mean residual of the two beams as discussed in the text. The boxed region spans the wavelength range where the Doppler-shifted H$_3^+$ line would be and has an RMS of 0.92\%.
\label{tauboo}}
\end{figure}

\begin{figure}
\includegraphics[width=\textwidth]{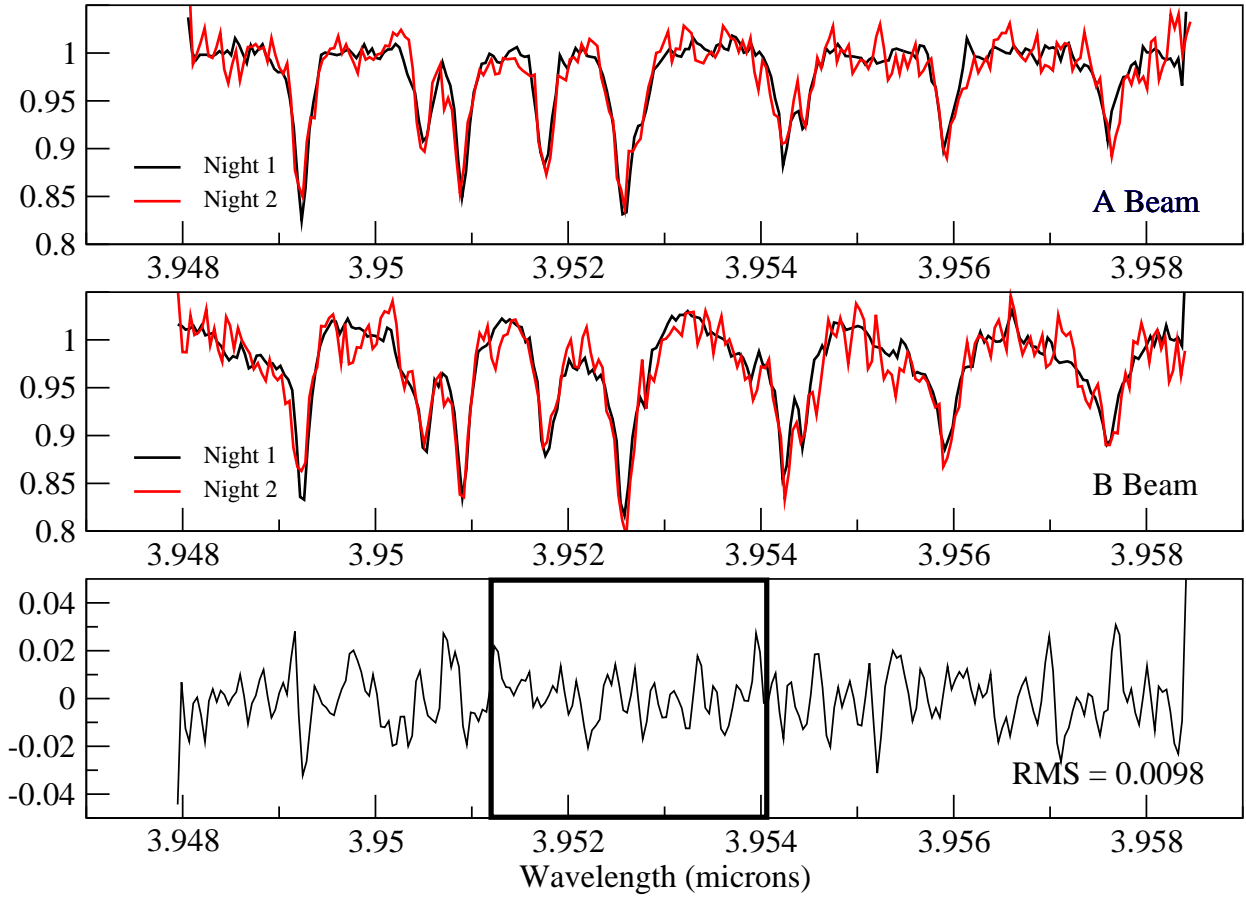}
\caption{Spectra and the mean residual of 55 Cnc. See caption to Figure~\ref{tauboo}.
\label{55Cnc}}
\end{figure}

\begin{figure}
\includegraphics[width=\textwidth]{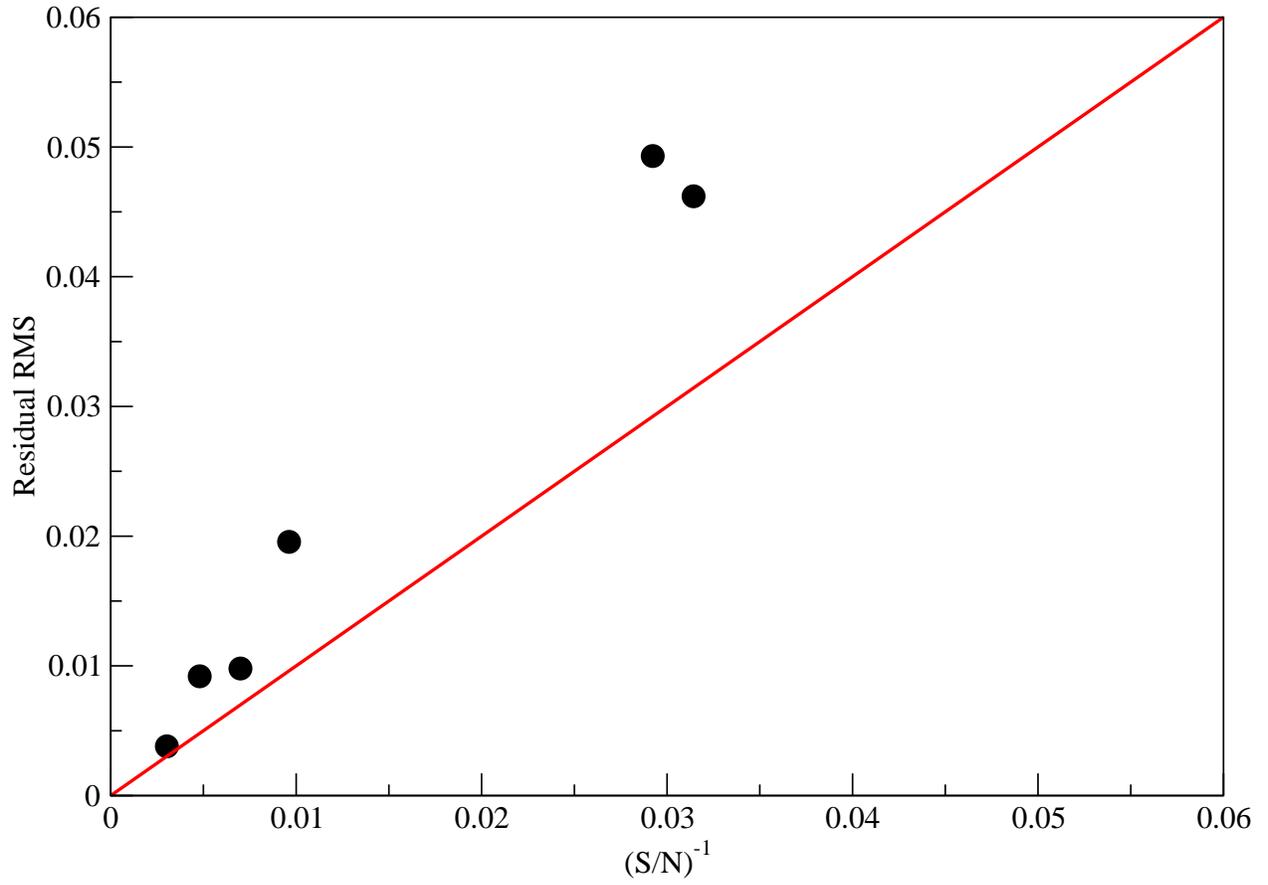}
\caption{Residual RMS plotted against (S/N)$^{-1}$ for each target star.  The red line represents the theortical S/N limit. $\upsilon$~And is the only star for which we reach this limit.\label{rms_SN}}
\end{figure}

\begin{figure}
\includegraphics[width=\textwidth]{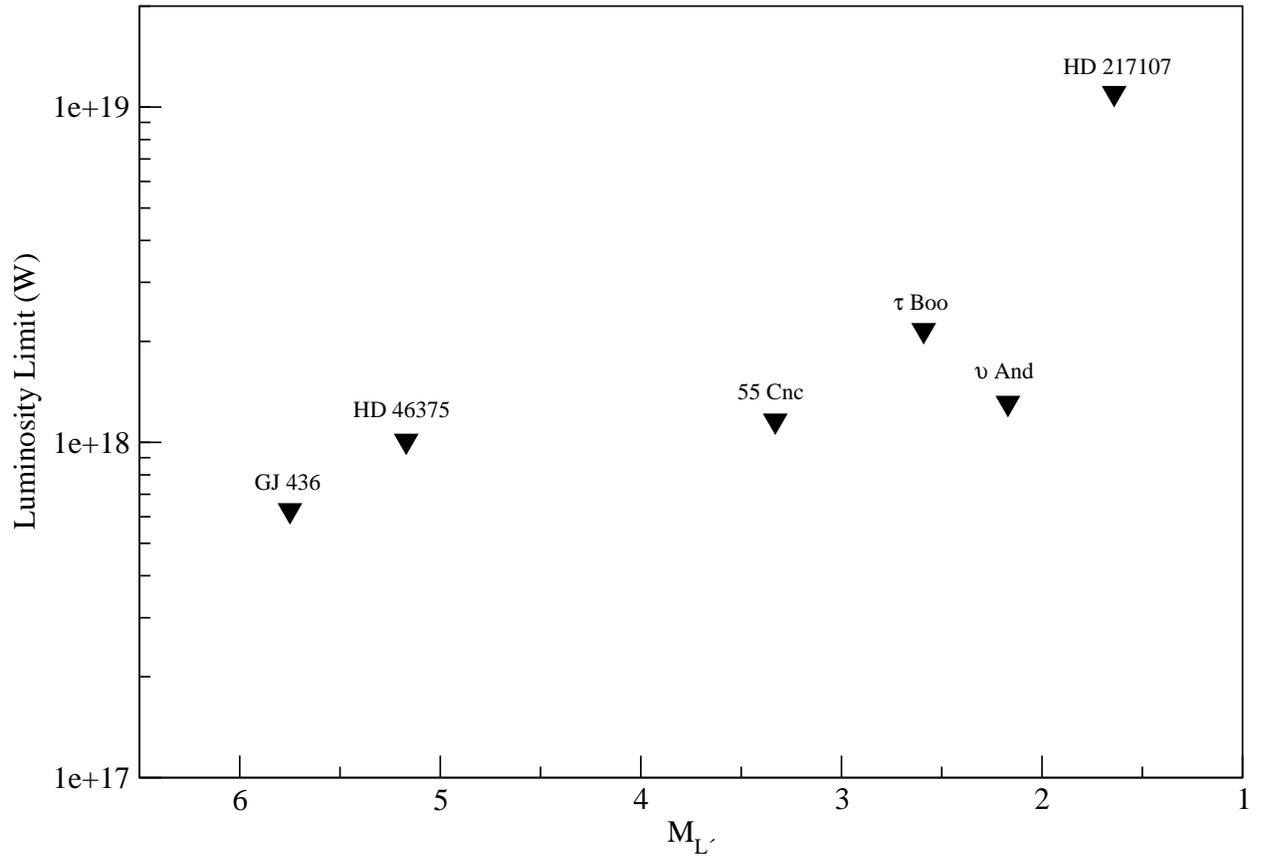}
\caption{Residual RMS converted into equivalent H$_3^+$ luminosity limit plotted against absolute L$^\prime$ of the star.  \label{limits}}
\end{figure}

The normalized spectra of $\tau$~Boo and 55 Cnc are presented in Figures~\ref{tauboo} and \ref{55Cnc}.  For these lower S/N spectra the RMS values are 0.92\% and 0.98\%, respectively, corresponding to luminosity limits of 2.2$\times$10$^{18}$ and 1.2$\times$10$^{18}$ W. In Figure~\ref{rms_SN}, we plot the RMS of the night-to-night residuals for the six targets against the theoretical noise limit, (S/N)$^{-1}$.  Observations of $\upsilon$~And reach this limit but as the S/N decreases, the RMS deviates from the line.  This is mainly due to systematic noise in the CSHELL detector, including the ``noisy" pixels, which are more difficult to identify and remove with lower S/N data. The upper limits for all the targets are listed in Table~\ref{stars} and plotted against absolute L$^\prime$ magnitude in Figure~\ref{limits}. Clearly, higher S/N data is not the sole contributor to a more stringent H$_3^+$ luminosity limit.  The lowest limit is set by the faintest and lowest S/N star because it is the closest target in our sample, GJ~436 (SpT = M2.5). GJ~436, at a distance of only 10.2 pc from Earth, has an H$_3^+$ emission limit of 6.3$\times$10$^{17}$ W, thus approaching Miller et al.'s (2000) estimate of $\sim$10$^{17}$~W.

\section{Summary and discussion\label{summary}}

We searched for the Q(1,0) transition of the H$_3^+$ molecule at 3.953~$\micron$ from the thermospheres of close-in giant planets around six stars and exploited the Doppler shift due to the planet's orbital motion to subtract the stellar photosphere and telluric lines.  The majority of our upper limits on H$_3^+$ luminosity are higher than predictions by Miller et al.~(2000) and Yelle (2004).  GJ~436, our faintest, nearest and lowest-S/N target, set the most stringent limit of 6.3$\times$10$^{17}$~W, comparable to Miller et al.'s most conservative estimate, but far higher than that of Yelle (2004). Our limits suggest that non-radiation effects (e.g.~magnetospheric heating) do not dramatically enhance H$_3^+$ emission from these planets.

The brighter (but more distant) stars in our survey offer the least promise for improvement. The limit on the H$_3^+$ emission from the planets orbiting $\upsilon$~And, our brightest target with the highest S/N spectra, is 1.3$\times$10$^{18}$~W. 
To achieve Miller et al.'s limit of $\sim$10$^{17}$~W, a S/N of 3400 pixel$^{-1}$ is necessary, and 10 times that amount to reach Yelle's limit. This would require a prohibitively long exposure time considering that we reached a S/N of only 300 in 2 hours of integration. 
We compare these requirements with the capabilities of NIRSPEC, the near-IR spectrograph mounted on the Keck II 10-m telescope:
A S/N of 3400 pixel$^{-1}$ at NIRSPEC's comparable spectral resolution requires 7 hours of integration time, but systematic effects would inhibit achieving such a high S/N.

The situation is better for less luminous but closer stars for which the contrast ratio between planetary H$_3^+$ emission and the stellar photosphere is higher.  The only such star in our sample, GJ~436, requires a relatively low S/N of $\sim$100 pixel$^{-1}$ to achieve a limit of 10$^{17}$~W. This can be reached in 10 hours with CSHELL and in only one hour with NIRSPEC. Yelle's limit of $\sim$10$^{16}$ W remains unfeasible even with Keck, the largest optical telescope in the world.

Yelle (2004) suggested that `occultation spectroscopy', as discussed by Richardson et al.~(2003), may be the best way to observe H$_3^+$ in planetary atmospheres. HD~209458~b (Charbonneau et al.~2000, Henry et al.~2000),  TrES-1 (Alonso et al.~2004) and HD 189733~b (Bouchy et al.~2005) are currently the only known planets for which this is practical.  
Given that our intranight residual RMS is half that of the RMS between nights, comparing spectra before, during and after secondary eclipse is an attractive option. However, one would not get the required S/N for these stars during the $\approx$2 hours of a single eclipse event, meaning cumulative data over several transits would be necessary.  A Jupiter-sized planet transiting a nearby M dwarf would be the most promising target for a future H$_3^+$ search, though such a planet has yet to be discovered.

\acknowledgments

We thank John Rayner, Alan Tokunaga, Tim Brown, and Sean Brittain  for helpful discussions. E.S. acknowledges the financial support of the NRC's Research Associateship Program.  E.G. acknowledges support by the NASA Terrestrial Planet Finder Foundation Science Program. This research has made use of the VizieR catalogue access tool, CDS, Strasbourg, France, and the Extrasolar Planet Catalog maintained by Jean Schneider. 



{\it Facilities:}\facility{ IRTF (CSHELL)}.

\clearpage

\begin{deluxetable}{cccccll}
\tabletypesize{\scriptsize}
\tablecaption{Planetary systems observed\label{planets}}
\tablewidth{0pt}
\tablehead{
\colhead{Star} & \colhead{Planet} & \colhead{Msin$i$} & \colhead{P$_{orb}$} & \colhead{$a$} & \colhead{Reference}  \\
\colhead{} & \colhead{} & \colhead{(M$_J$)} & \colhead{(days)} & \colhead{(AU)} & \colhead{} 
}
\startdata

GJ 436 		& b      & 0.07  & 2.64  & 0.03 & Butler et al.~2004\\
55 Cnc  		&e   &0.045 & 2.81 & 0.038 & McArthur et al.~2004\\
        			&b      &0.784  &14.67  &0.115  & Butler et al.~1997\\
       			&c      &0.217  &43.93  &0.24   & Marcy et al.~2002\\
        			&d      &3.92   &4517.4 &5.257  & Marcy et al.~2002\\
$\upsilon$ And  	 	&b      &0.69   &4.617  &0.059 & Butler et al.~1997\\
        			&c      &1.89   &241.5  &0.829  & Butler et al.~1999\\
       		 	&d      &3.75   &1284   &2.53   & Butler et al.~1999\\
$\tau$ Boo 		&b      &4.13   &3.3135 &0.046  & Butler et al.~1997\\
HD 46375      &b      &0.249  &3.024  & 0.041 & Marcy et al.~2000\\
HD 217107    &b      &1.37   &7.1269 & 0.074 & Fischer et al.~1999\\
        			&c      &2.1    &3150   &4.3 & Vogt et al.~2000\\
~                                                
\enddata
\end{deluxetable}

\clearpage
\begin{deluxetable}{ccccccccccccrl}
\tabletypesize{\scriptsize}
\tablecaption{Targets and observations\label{stars}}
\tablewidth{0pt}
\tablehead{
\colhead{Star} & \colhead{SpT} & \colhead{Dist.} & \colhead{K\tablenotemark{a}} & \colhead{L$^{\prime}$\tablenotemark{b}} & \colhead{HJD - Night 1} & \colhead{Total Exp.} & \colhead{Total S/N} &\colhead{RMS of} & \colhead{Emission} \\
\colhead{} & \colhead{} & \colhead{(pc)} & \colhead{} & \colhead{} & \colhead{HJD - Night 2} & \colhead{Time (s)} & \colhead{(pixel$^{-1}$)} & \colhead{Residual} & \colhead{Limit (W)} 
}
\startdata

 GJ 436	&	         M2.5     	&	10.2	&	6.07	&	5.80	&	2453417.97	&	6480	&	30	&	0.0493	&	6.3E+17	\\
	&		&		&		&		&	2453418.98	&	5760	&	40	&		&		\\
 55 Cnc 	&	         G8 V     	&	13.4	&	4.02		&	3.96		&	2453417.87	&	6360	&	310	&	0.0098	&	1.2E+18	\\
	&		&		&		&		&	2453419.85	&	960	&	100	&		&		\\
$\upsilon$ And  	&	         F8 V     	&	13.47	&	2.86 	&	2.82		&	2453610.10 &	7280	&	310 	&	0.0038	&	1.3E+18	\\
	&		&		&		&		&	2453612.06	&	7520	&	370 	&		&		\\
$\tau$ Boo	&	         F7 V     	&	15	&	3.51		&	3.47		&	2453418.06	&	3840	&	270	&	0.0092	&	2.2E+18	\\
	&		&		&		&		&	2453420.05	&	2520	&	180	&		&		\\
HD 46375  	&	        K1 IV     	&	33.4	&	7.85		&	7.79		&	2453417.81	&	2640	&	30	&	0.0462	&	1.0E+18	\\
	&		&		&		&		&	2453419.75	&	7440	&	30	&		&		\\
HD 217107  	&	        G8 IV     	&	37	&	4.54		&	4.49		&	2453609.94	&	4320	&	160	&	0.0196	&	1.1E+19	\\
	&		&		&		&		&	2453611.91	&	4080	&	80	&		&		\\
\enddata
\tablenotetext{a}{K magnitudes taken from the Two Micron All Sky Survey (2MASS) via VizieR.}
\tablenotetext{b}{Calculated from K magnitudes and intrinsic colors from Bessell \& Brett (1988).}
\end{deluxetable}

\clearpage

\clearpage


\clearpage

\clearpage
\clearpage
\clearpage

\end{document}